\documentclass[a5paper,portuguese,twoside,10pt]{article}
\usepackage[a5paper]{geometry}

\newcommand{\titulo}[1]{\Large{\textbf{#1}\vspace*{.2cm}} }
\newcommand{\autor}[2]{\normalsize{\textsf{#1 #2} \index{\textbf{#2}, #1}}}
\newcommand{\afil}[2]{\mbox{} \\ #1, \textit{#2} \vspace*{.1cm} \\}
\newcommand{\pchave}[1]{{\textbf{Palavras--chave}: #1\\}\vspace{-0.2cm}}

\newcommand\halmos{\rule{0.1in}{0.1in}}

\newcommand{\bibliografia}[1]{{\small \begin{thebibliography}{99} #1 \end{thebibliography}}}

\usepackage[portuguese]{babel}
\usepackage{ae}
\usepackage{graphicx}
\usepackage{epstopdf}
\usepackage{amssymb,amsmath}
\usepackage{bm}
\usepackage[utf8]{inputenc}
\usepackage{imakeidx}
\makeindex[title={\'{I}ndice de Autores},columns=2]

\usepackage{eurosym}
\usepackage{icomma}
\usepackage{url}
\usepackage{color}
\usepackage{subfigure}

\usepackage{fancyhdr}
\usepackage{amssymb}
\usepackage{wasysym}
\usepackage[usenames,dvipsnames]{xcolor}

\usepackage{multicol}
\usepackage{booktabs}
\usepackage{multirow}
\usepackage{caption}

\AtBeginDocument{%

        \renewcommand{\pchave}{\textbf{Keywords}: }
        }
\begin{document}

\titulo{Threshold selection for wave heights: \\ asymptotic methods based on L-moments}


\autor{Jessica}{Silva Lomba}
\afil{CEAUL -- Centro de Estatística e Aplicações, Faculdade de Ciências, Universidade de Lisboa, Portugal}{jslomba@fc.ul.pt}

\autor{Maria Isabel}{Fraga Alves}
\afil{CEAUL -- Centro de Estatística e Aplicações, Faculdade de Ciências, Universidade de Lisboa, Portugal}{isabel.alves@fc.ul.pt}

\pchave{Extreme Values; Threshold Selection; L-moments}

\medskip

\textbf{Abstract}: Two automatic threshold selection (TS) methods for Extreme Value analysis under a peaks-over-threshold (POT) approach are presented and evaluated, both built on: fitting the Generalized Pareto distribution (GPd) to excesses' samples over candidate levels; the GPd-specific relation between L-skewness and L-kurtosis; the asymptotic behaviour of the matching L-statistics. Performance is illustrated on significant wave heights data sets and compared to the L-moment-based heuristic in \cite{lombaalves2020}, which is found to be favorable.

\section{Introduction}
The POT approach in Extreme Value analysis consists in selecting observations that fall above a pre-specified high threshold $u$ and fitting a GPd to the excesses of this level, under the Pickands-Balkema- de Haan Theorem (\cite{BalkDeHaan1974}, \cite{pickands1975}). In several applied  fields, risk analysis can be built on inference of high quantiles -- also known as return levels (RL) -- based on this fit, its accuracy thus highly dependent on an adequate choice of $u$. The problem of choosing $u$ is an open topic (see \cite{scarrottmacdonald2012} and \cite{langousis2016} for recent reviews) and lies on balancing bias \emph{vs.} variance in estimation resulting from thresholds that are too low (poor fit) or too high (few excesses).  

Addressing usual issues in selection approaches, \cite{lombaalves2020} suggested the ALRSM, a heuristic automatic selection method based on L-moments and the L-moment Ratio Diagram, shown to eliminate subjectivity while enjoying accuracy and efficiency for small and large samples. However, it provides no theoretical assurance on the quality of the GPd fit above the selected $u$: a level is always chosen despite poor fitting. In this work we present two asymptotically supported methodologies that aim to meet the same standards of the ALRSM, while evaluating adequacy of the fit and consequently of the selected level.

\subsection{L-moments and L-statistics}
Hosking \cite{hosking1986} introduced the \emph{L-moments} as specific linear combinations of Probability Weighted Moments (PWM) that can be read as measures of location, scale and shape of distributions, allowing for its easy description, identification and estimation of parameters. The full set of L-moments of a random variable $X$ with distribution function $F$ exists if $\mathbb{E}|X|<\infty$, a broader set-up than that for conventional moments. Given the PWM $\alpha_r=M_{1,0,r}=\mathbb{E}\left[X\{1-F(X)\}^r\right]$, the first four L-moments are
\begin{equation}\label{lmom}\small
\begin{gathered}
    \lambda_1 = \alpha_0\,\text{(expected value)}\quad ,\quad \lambda_2 = \alpha_0-2\alpha_1\,\text{(L-scale)},\\ 
    \lambda_3 = \alpha_0-6\alpha_1+6\alpha_2\quad \text{and} \quad \lambda_4 = \alpha_0-12\alpha_1+30\alpha_2-20\alpha_3\,.
\end{gathered}\normalsize
\end{equation}It is also useful to define scale-independent L-moment ratios, of which we will use $\tau_3 = \frac{\lambda_3}{\lambda_2}$ and $\tau_4 = \frac{\lambda_4}{\lambda_2}$, known resp. as L-skewness and L-kurtosis, globally bounded as
\begin{equation}\label{genbounds}\small
    \frac{1}{4} \left( 5\,\tau_3^2-1\right )\leq\tau_4 < 1\,.
\end{equation}\normalsize
In particular for the GPd of scale and shape (tail weight) parameters {\small$(\sigma_u,\,\xi)\in\mathbb{R}^{+}\times\mathbb{R}$} --  \scriptsize{d.f. $\left(1-\left[1+\tfrac{\xi}{\sigma_u}x\right]_{+}^{-1/\xi}\right)$, $x>0$, $a_+=\max(a,0)$} \normalsize-- we have\vspace{-0.2cm}\small
\begin{alignat}{3}
 &\lambda_1=\dfrac{\sigma_u}{1-\xi}\;, && \qquad \lambda_2=\dfrac{\sigma_u}{(1-\xi)(2-\xi)}\;, &&
    \qquad \tau_3=\dfrac{1+\xi}{3-\xi}\;,
    \label{lambda1e2etau3GP}
\end{alignat}
\normalsize defined for $\xi<1$, along with the following specific relationship between L-skewness and L-kurtosis:

\begin{equation}\label{relationtauGP}\small
    \tau_4=\tau_3\,\frac{1+5\tau_3}{5+\tau_3}=:g(\tau_3)\;.
\end{equation}\normalsize

Corresponding estimators -- \emph{L-statistics} -- are simply found as linear combinations of the ordered observations $x_{1:n}\leq \ldots \leq x_{n:n}$, with several theoretical advantages over conventional sample moments \cite{hoskingwallis1997}. The unbiased L-statistics of interest $(\ell_1,\,\ell_2,\,\ell_3,\,\ell_4)$ are found by replacing the PWM $\alpha_r$ in \eqref{lmom} by their unbiased estimators \vspace{-0.cm}
\begin{equation}\label{a_r}\small  a_r=\frac{1}{n}\,\sum_{i=1}^{n}\binom{n-i}{r}\,x_{i:n}\,\binom{n-1}{r}^{-1}\,,\qquad r=0,1,\ldots,n-1.
\end{equation} \normalsize

\vspace{-0.cm}The ratios statistics -- here $\mathnormal{t}_3 = \frac{\ell_3}{\ell_2}$ and $\mathnormal{t}_4 = \frac{\ell_4}{\ell_2}$ -- are only asymptotically unbiased. 

Useful estimators of the GPd parameters follow from \eqref{lambda1e2etau3GP} and \eqref{a_r} as \small
\vspace{-0.1cm}\begin{alignat}{2} \label{PWMxisigma}
 &\widehat{\xi} = 2-\frac{\ell_1}{\ell_2} = 2-\frac{a_0}{a_0-2a_1}\,, \qquad
 &\widehat{\sigma}_u = \ell_1 \left(1-\widehat{\xi}\right) = \frac{2a_0\,a_1}{a_0-2a_1}\,.
\end{alignat}\normalsize 

\vspace{-0.05cm}If, additionally, var$[X]<\infty$, it is possible to demonstrate the asymptotic normality of $a_r$, $\ell_r$ and $\mathnormal{t}_r$, as well as compute the corresponding asymptotic bias and variance. It has been empirically shown that, in small samples, these estimators approximate their asymptotic normality more closely than traditional sample moments, often closely enough for samples as small as $n=20$ \cite{hosking1986}. 

Considering again the GPd, with restricted shape $\xi \in \left(-\frac{1}{2},\frac{1}{2}\right)$ (finite variance), we have that the $a_r$ are asymptotically Normal with 
 \begin{equation}\label{Ars}\small
         \boldsymbol{\mathnormal{A}_{r,s}} = \lim_{n \to \infty} n\,\text{cov}(a_r,a_s) =  \frac{\sigma_u^2}{(r+1-\xi)(s+1-\xi)(r+s+1-2\,\xi)},
\end{equation} \normalsize  

\vspace{-0.005cm}$r,s=0,1,\ldots,n-1$. 
These $\mathnormal{A}_{r,s}$ give us the asymptotic var-covariances of $\ell_r$ and $\mathnormal{t}_r$: we compute matrix 
$\boldsymbol{\Lambda} := \lim_{n \to \infty} n\,\mathbb{V}\text{ar}(\ell_1,\ell_2,\ell_3,\ell_4) = M\,\boldsymbol{\mathnormal{A}}\,M^{T}$ ($M$ is a $4\times4$ 
numeric matrix given in \cite{hosking1986}) and matrix $\boldsymbol{\mathnormal{T}} := \lim_{n \to \infty} n\,\mathbb{V}\text{ar}(t_3,t_4)$ for the asymptotic bi-Normal distribution of $(\mathnormal{t}_3,\mathnormal{t}_4)$, with entries

\small\begin{align*}
&\boldsymbol{T_{33}} = \lim_{n \to \infty} n\,\text{var}(t_3) =
    \frac{\Lambda_{33}-2\,\tau_{3}\,\Lambda_{23}+\tau_{3}^2\,\Lambda_{22}}{\lambda_2^2} \nonumber
\end{align*}
\noindent\noindent\begin{align}\label{Tentries}
&\boldsymbol{T_{34}} =\lim_{n \to \infty} n\,\text{cov}(t_3,t_4) =
    \frac{\Lambda_{34}-\tau_{3}\,\Lambda_{24}-\tau_{4}\,\Lambda_{23}+\tau_{3}\,\tau_{4}\Lambda_{22}}
    {\lambda_2^2}\\
&\boldsymbol{T_{44}} =\lim_{n \to \infty} n\,\text{var}(t_4) =
    \frac{\Lambda_{44}-2\,\tau_{4}\,\Lambda_{24}+\tau_{4}^2\,\Lambda_{22}}{\lambda_2^2}\,.\nonumber
\end{align}
\normalsize

\vspace{-0.1cm} All entries of $\boldsymbol{\mathnormal{A}}$, $\boldsymbol{\Lambda}$ and $\boldsymbol{T}$ can be estimated by plugging-in the sample PWM $a_0$ through $a_3$, in order to obtain $\widehat{\xi}$, $\widehat{\sigma}_u$, $\widehat{\lambda}_2\equiv\ell_2$, $\widehat{\tau_3}\equiv t_3$, and $\widehat{\tau_4}\equiv t_4$, thus yielding the PWM-based estimates $\boldsymbol{\widehat{\mathnormal{A}}}$, $\boldsymbol{\widehat{\mathnormal{\Lambda}}}$ and $\boldsymbol{\widehat{\mathnormal{T}}}$.

\subsection{L-moment Ratio Diagram (LMRD)}
\begin{figure}[h!]
    \centering\vspace{-0.2cm}
    \includegraphics[width=0.5\textwidth]{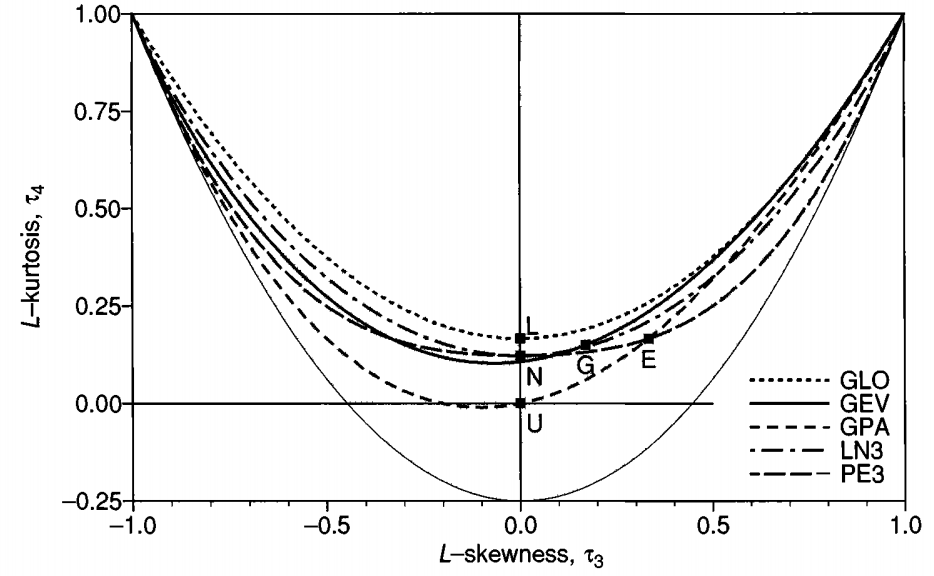}
    \vspace{-0.2cm}\caption{L-moment ratio diagram -- Figure 2.5, page 25 of \cite{hoskingwallis1997}.}
    \label{fig:lmrd}
\end{figure}
The LMRD commonly refers to the representation of the ratios $\tau_3$ and $\tau_4$ for several distributions, a visualization of L-kurtosis \textit{vs.} L-skewness, used in Regional Frequency Analysis for regional distribution choice \cite{hoskingwallis1997}. As shown in Figure \ref{fig:lmrd}, three-parameter distributions plot as a line, with different values of the shape parameter corresponding to different points on that line; distributions with more than one shape parameter can comprehend a 2-dimensional region. For the GPd, a specific shape-$\xi$ corresponds to a single point $(\tau_3,\tau_4)=\left(\tau_3,g(\tau_3)\right)$ on the curve given by \eqref{relationtauGP}, with negative values of $\tau_3$ relating to $\xi<-1$, very uncommon extremely light tails.

Agreement of a sample to a distribution can be judged by the proximity between $(\mathnormal{t}_3,\,\mathnormal{t}_4)$ and the theoretical curve of interest. This is the main concept on which the ALRSM is based, and one we explore in the sequel, now minding the asymptotic considerations above.

\section{Automatic Threshold Choice}
\subsection{Automatic L-moment Confidence Band Selection Method (ALCBSM)}


The consistent estimators $(\mathnormal{t}_3,\,\mathnormal{t}_4)$, under the GPd, asymptotically follow a bi-Normal distribution with var-covar matrix given by $\boldsymbol{T}$ in \eqref{Tentries}: $\sqrt{n}\left[(\mathnormal{t}_3-\tau_3)\quad(\mathnormal{t}_4-\tau_4) \right]^T\xrightarrow{d}\boldsymbol{\mathcal{N}}\left(\boldsymbol{0}, \boldsymbol{T}\right)$ (c.f. \cite{hosking1986}).
Noting that the conditional distribution of jointly Normal variables is still Gaussian, we can study the asymptotic behaviour of the sample L-kurtosis given an estimate of the L-skewness, and \textit{vice-versa}. As such, we will build our LMRD confidence bands on the fact that
\vspace{-0.1cm}\begin{equation}\label{distt4dadot3}\small
    \mathnormal{t}_4\,|\,\mathnormal{t}_3=t^{*} \overset{a}{\frown}\mathcal{N}\left(\mathbb{E}\left(\mathnormal{t}_4\,|\,\mathnormal{t}_3=t^{*}\right), \text{var}\left(\mathnormal{t}_4\,|\,\mathnormal{t}_3=t^{*}\right)\right)\,\vspace{-0.3cm}
\end{equation}
\small\begin{align*}
    &\text{with } \mathbb{E}\left(\mathnormal{t}_4\,| \,\mathnormal{t}_3=t^{*}\right)=\mathbb{E}\left(\mathnormal{t}_4\right) + \frac{\text{cov}(t_3,t_4)}{\text{var}(t_3)}\left(t^{*}-\mathbb{E}\left(\mathnormal{t}_3\right)\right)\sim \tau_4 + \frac{\boldsymbol{T_{34}}}{\boldsymbol{T_{33}}}\left(t^{*}-\tau_3\right)\\
    &\text{and }\text{var}\left(\mathnormal{t}_4\,| \,\mathnormal{t}_3=t^{*}\right)=\text{var}(t_4).(1-\rho^2)\sim\frac{\boldsymbol{T_{44}}}{n}\left(1-\frac{\boldsymbol{T_{34}}^2}{\boldsymbol{T_{33}}\,\boldsymbol{T_{44}}}\right).
\end{align*}

\vspace{-0.1cm}
\normalsize In practice, true L-skewness and L-kurtosis are unknown: we plug-in the estimates $\widehat{\tau_3}\equiv\mathnormal{t}_3=t^{*}$, and $\widehat{\tau_4}= g(\mathnormal{t}_3)=g(t^{*})$, with $g(.)$ in \eqref{relationtauGP}. Thus, the expectation of the conditional in \eqref{distt4dadot3} is reduced to the parameter of interest, $\tau_4$, allowing us to devise confidence intervals (CI's) for its value. 
As such, given an observed $t^*$ of $\mathnormal{t}_3$, from a sample of size $n$, we estimate with approx. $(1-\alpha)\%$ confidence that \small{\begin{align}\label{ICt4dadot3}
    &\tau_4\in\left[LCI_{\tau_4}\,;\,UCI_{\tau_4}\right] =  \\
    &\left[g(t^*)-z_{1-\frac{\alpha}{2}}\sqrt{\frac{\widehat{\boldsymbol{T_{44}}}}{n}\left(1-\widehat{\boldsymbol{\rho_{34}}}^2\right)}\,;\, g(t^*)+z_{1-\frac{\alpha}{2}}\sqrt{\frac{\widehat{\boldsymbol{T_{44}}}}{n}\left(1-\widehat{\boldsymbol{\rho_{34}}}^2\right)}\,\right]\nonumber
\end{align}}\normalsize
where $\widehat{\rho_{34}}^2=\frac{\widehat{\boldsymbol{T_{34}}}^2}{\widehat{\boldsymbol{T_{33}}}\,\widehat{\boldsymbol{T_{44}}}}$ and $z_{1-\frac{\alpha}{2}}$ is the \small{$\left(1-\frac{\alpha}{2}\right)$}\normalsize-probability quantile of the standard Normal.

\vspace{0.05cm}If we instead consider the distribution of $\mathnormal{t}_3\,|\,\mathnormal{t}_4=t^{*}$, the results are perfectly analogous and therefore their derivation is here overlooked: given an observed  $t^*$ of $\mathnormal{t}_4$, from a sample of size $n$, we estimate with approx. $(1-\alpha)\%$ confidence that \vspace{-0.1cm}
\small{\begin{align}\label{ICt3dadot4}
    &\tau_3\in \left[LCI_{\tau_3}\,;\,UCI_{\tau_3}\right] = \\
    &\left[g^{-1}(t^*)-z_{1-\frac{\alpha}{2}}\sqrt{\frac{\widehat{\boldsymbol{T_{33}}}}{n}\left(1-\widehat{\boldsymbol{\rho_{34}}}^2\right)}\,;\, g^{-1}(t^*)+z_{1-\frac{\alpha}{2}}\sqrt{\frac{\widehat{\boldsymbol{T_{33}}}}{n}\left(1-\widehat{\boldsymbol{\rho_{34}}}^2\right)}\,\right] \nonumber \vspace{-0.1cm}
\end{align}}\normalsize
where $g^{-1}(\tau_4):=\frac{\tau_4-1}{10}+\frac{1}{10}\sqrt{\tau_4^2+98\tau_4+1}$ is the analytical inverse of $g(.)$, restricted to the 1\textsuperscript{st} quadrant of the LMRD ($\tau_3,\tau_4>0$). 

These CI's can be used for evaluating the acceptability of the GPd fit to the sample. By estimating the L-kurtosis in \eqref{ICt4dadot3} as $g(t_3)$, rather than its PWM-based estimator $t_4$, we get a CI for $\tau_4$ that is centered around the GPd curve in the LMRD \eqref{relationtauGP}, as well as completely independent from the estimate $t_4$ itself. As such, we can judge a sample to be sufficiently well adjusted by the GPd if, given the estimate $t_3$, the sample-computed $t_4$ falls within the corresponding bounds of the estimated CI \eqref{ICt4dadot3}. A similar reasoning is valid when exchanging $t_3$ by $t_4$ and using the interval in \eqref{ICt3dadot4}.

Return to the threshold selection setting: in the POT-GP approach, given i.i.d. data from an unknown distribution (assumed to belong to some max-domain of attraction -- c.f. \cite{BalkDeHaan1974}, \cite{pickands1975}), we have to choose, from a reasonable set of candidates, a level $u^*$ after which the GP approximation to the sample of excesses is judged to hold sufficiently well. In the present framework, $u^*$ will be \textbf{automatically} chosen as the lowest candidate for which $(\mathnormal{t}_3,\mathnormal{t}_4)$, computed from the $n_{u*}$ excesses, simultaneously fall inside the respective CI \eqref{ICt4dadot3} and \eqref{ICt3dadot4}, computed in turn by fixing first the value of $\mathnormal{t}_3$ and then that of $\mathnormal{t}_4$.

\vspace{0.15cm}$\xrightarrow{}$ Given a sample $x_1,\ldots,\,x_n$ of size $n$ and $\left\{u_i\right\}_{i=1}^I$ a reasonable set of candidate thresholds (we suggest $I=10$ or $I=20$ equal-step sample quantiles, starting at 25\%), the ALCBSM works as follows:
\begin{enumerate}
    \item For each candidate threshold $u_i$, $i=1,\ldots,I$:
    \small{\begin{itemize}
        \item[a)]  Compute the sample L-skewness and L-kurtosis for the excesses over each candidate $(\mathnormal{t}_{3,u_i},\mathnormal{t}_{4,u_i})$, as well as the GPd-specific functions $g({t}_{3,u_i})$ and $g^{-1}({t}_{4,u_i})$;
        
        \item[b)] Compute the estimates of the parameters $\xi$ and $\sigma_u$ in \eqref{PWMxisigma} and plug them into \eqref{Ars} to obtain $\widehat{\boldsymbol{A}}$ and $\widehat{\boldsymbol{\Lambda}}$;
        
        \item[c)] Compute the bounds $LIC_{\tau_4}^i$ and $UIC_{\tau_4}^i$ in \eqref{ICt4dadot3} by plugging $\ell_2$, $\widehat{\boldsymbol{\Lambda}}$, $\widehat{\tau_3}\equiv {t}_{3,u_i}$ and $\widehat{\tau_4}=g({t}_{3,u_i})$ into \eqref{Tentries};
        
        \item[d)] If ${t}_{4,u_i}$ falls within the values of $LIC_{\tau_4}^i$ and $UIC_{\tau_4}^i$, proceed to step 1.e); otherwise, update the candidate threshold to $u_{i+1}$ and return to step 1.a);
        
        \item[e)] Compute the bounds $LIC_{\tau_3}^i$ and $UIC_{\tau_3}^i$ in \eqref{ICt3dadot4} by plugging $\ell_2$, $\widehat{\boldsymbol{\Lambda}}$, $\widehat{\tau_4}\equiv {t}_{4,u_i}$ and $\widehat{\tau_3}=g^{-1}({t}_{4,u_i})$ into \eqref{Tentries};
        
        \item[d)] If ${t}_{3,u_i}$ falls within the values of $LIC_{\tau_3}^i$ and $UIC_{\tau_3}^i$, proceed to step 2; otherwise, update the candidate threshold to $u_{i+1}$ and return to step 1.a);
    \end{itemize}}\normalsize
   
    \item \normalsize The lowest threshold above which the underlying distribution's tail behaviour can be considered approximately GPd is automatically selected as $u^*=u_{i}$ -- the first level above which the corresponding L-statistics fall close enough to the curve, inside the $(1-\alpha)\%$ confidence bands. No threshold is selected if no pair $(\mathnormal{t}_{3,u_i},\mathnormal{t}_{4,u_i})$ simultaneously falls inside both CI's.
\end{enumerate}

\subsection{Automatic L-moment Goodness-of-Fit Selection Method (ALGFSM)}

\normalsize For the development of the alternative methodology we now present, two main techniques were combined: computation of the \emph{goodness-of-fit} (GoF) \emph{measure} suggested in Chapter 5 of \cite{hoskingwallis1997}, together with the \emph{ForwardStop} stopping rule used for automatic selection by \cite{baderetal2018}.  Not unlike our previous framework, the way to test the quality of the GPd fit to a given sample of excesses will be based on the behaviour of $(t_3,\,t_4)$ in regard to $(\tau_3,\,\tau_4)$. We will again make use of the asymptotic Normality of these estimators, but another approach is taken for estimation of the corresponding variability -- simulation.

Let us introduce the four-parameter Kappa distribution (c.f. \cite{hosking1994}), given as function of $\mu\in\mathbb{R}$, $\sigma>0$, $\xi\in\mathbb{R}$ and $h\in\mathbb{R}$, respectively location, scale and two shape parameters. The Kappa family counts as special cases the GPd ($h=1$), the Generalized Extreme Value distribution (GEVd, $h=0$) and other distributions of interest in various fields. Hence, it is convenient when commitment to one such specific behaviour is not desired, or when a simpler two/three-parameter distribution does not provide a sufficiently accurate fit. The cumulative distribution function is given as\vspace{-0.3cm} \small

\begin{equation}\label{Kappa}
  \text{Kappa}(x|\mu,\,\sigma,\,\xi,\,h):=\left\{
  \begin{aligned}
   &\left(1-h\,\left[1+ \xi\,\frac{x-\mu}{\sigma}\right]^{-\frac{1}{\xi}}\right)^{\frac{1}{h}}, & \hspace{-10pt}h,\,\xi\neq0 \\
   &\text{GEV}(x|\mu,\,\sigma,\,\xi), & \hspace{-10pt}h=0,\,\xi\neq0 \\
   &\left(1-h\,\exp\left[-\frac{x-\mu}{\sigma}\right]\right)^{\frac{1}{h}}, & \hspace{-10pt}h\neq0,\,\xi=0 \\
   &\text{Gumbel}(x|\mu,\,\sigma), & \hspace{-10pt}h,\,\xi=0  \\
 \end{aligned}\right.\end{equation}\vspace{-0.1cm}\normalsize 
with the variable support being bounded above by $\mu-\frac{\sigma}{\xi}$ if $\xi<0$, and bounded below by $\mu + \frac{\sigma}{\xi}\left(1-h^{\xi}\right)$ if $h>0$,  or by $\mu + \frac{\sigma}{\xi}$ if $h\leq0$ and $\xi>0$ (with otherwise infinite right/left endpoints).

\begin{figure}[ht!]
  \centering
    \includegraphics[width=0.5\textwidth]{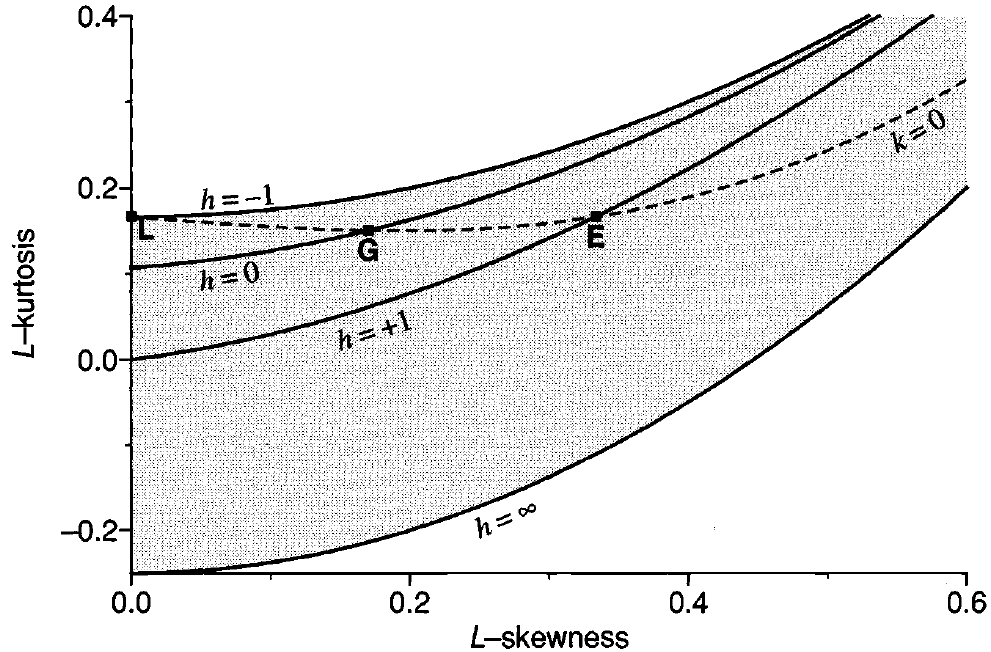}
    \vspace{-0.2cm}
    \caption{LMRD for the Kappa distr. -- Figure A.1, page 204 of \cite{hoskingwallis1997}. 
    }
    \label{fig:lmomratiodiagramKappa}
\end{figure}

For this distribution, the L-skewness and L-kurtosis are dependent on $\xi$ and $h$, and as such, in the LMRD, the possible $(\tau_3,\tau_4)$ pairs appear as a 2-dimensional sub-area of this plot, shown in Figure \ref{fig:lmomratiodiagramKappa}. 
The most useful range of parameters ($h\geq-1$) corresponds to the region between the line corresponding to the Generalized Logistic distribution ($h=-1$) and the general lower bound for all distributions \eqref{genbounds} -- the shaded area in Figure \ref{fig:lmomratiodiagramKappa}. Although analytical expressions for the parameters in terms of the L-moments do not exist, this problem can be circumvented using Newton-Raphson numerical methods.

The versatility of the Kappa distribution makes it suitable for checking robustness of statistical procedures under distributional assumptions that may not be verified -- this is the principle on which \cite{hoskingwallis1997} based their GoF test, evaluating the quality of adjustment to several candidate distributions. In our setting, the interest is in deciding after which candidate level are the excesses in a sample sufficiently well modeled by the GPd. Since this is a special case of the Kappa -- meaning that the simpler distribution can be used if appropriate -- we follow the process of artificial data generation in \cite{hoskingwallis1997}, to compare the observed and expected behaviours of the L-statistics. 

For a sample of $n_u$ excesses of a threshold $u$, compute the corresponding GPd-L-moment based GoF statistic $Z^{GP}_u$ as follows:\small{
\begin{itemize}
    \item Fit a GPd to the data using the method of L-moments -- since it is a three-parameter distribution (with null location), it is fitted with resource to $\ell_{1,u}$, $\ell_{2,u}$ and $t_{3,u}$, and as such the fitted distribution has L-skewness equal to the estimated $t_{3,u}$;
    \item Compute the theoretical L-kurtosis of the fitted GPd $\tau_{4,u}^{GP}$ -- the point in the LMRD curve \eqref{relationtauGP} corresponding to the abscissa $t_{3,u}$;
    \item Fit a Kappa distribution using the method of L-moments -- since it is a four-parameter distribution, it is fitted with resource to $\ell_{1,u}$, $\ell_{2,u}$, $t_{3,u}$ and $t_{4,u}$;
    \item Simulate a large number $N$ of samples of size $n_u$ from the fitted Kappa (\cite{hoskingwallis1997} suggest $N=500$) -- these provide estimates of bias $B_{4,u}$ and variability $\sigma_{4,u}$ of the $t_4$ for samples from this model;
    \item Compute the goodness-of-fit measure as
    \begin{equation}\label{ZGP}
        z^{GP}_u=\frac{\tau_{4,u}^{GP}-t_{4,u}+B_{4,u}}{\sigma_{4,u}}\,.
    \end{equation}
\end{itemize}}

\normalsize \vspace{-0.1cm}The $Z_u^{GP}$ can be considered approximately standard Normal. Thus, we judge the GPd fit to the excesses' sample sufficiently accurate if $|z_u^{GP}|\leq z_{1-\frac{\alpha}{2}}$ (with \footnotesize$z_{1-\frac{\alpha}{2}}$\normalsize the standard normal \footnotesize{$\left(1-\frac{\alpha}{2}\right)$}\normalsize-probability quantile, $\alpha$ the significance level -- \cite{hoskingwallis1997} suggest \small$\alpha=0.1$\normalsize). However, with the aim of choosing from an ordered set of candidate levels $\left\{u_i\right\}_{i=1}^I$, it is best to quantify acceptability of the adjustment through the \textit{p}-value of the corresponding double-sided Normality test \small\vspace{-0.2cm}
 \begin{equation}\label{pvalZGP}
 p_{u}=2-2\,\Phi\left(|z^{GP}_{u}|\right),\vspace{-0.1cm}
 \end{equation}\normalsize
with $\Phi(.)$ the the standard Gaussian distribution function.

Repeating this process for all candidate thresholds, we obtain the set of \textit{p}-values corresponding to the set of ordered hypothesis\vspace{-0.15cm}
\begin{equation*}
    H_0^i:\text{\emph{ the distribution of the }} n_i \text{\emph{ excesses above }} u_i \text{\emph{ follows the GPd,}}\vspace{-0.15cm}
\end{equation*}
for $i=1,\ldots,I$, the same context under which \cite{baderetal2018} establish their automatic selection procedure. So we make use of the same rejection rule as in the reference, the \emph{ForwardStop}, which under the (here violated) assumption of independence of the multiple tests, allows for control of the False Discovery Rate at a pre-set level $\alpha$. Based on the transformed sequence of \textit{p}-values obtained from the GoF measure $Z^{GP}$, the \emph{ForwardStop} consists of finding\vspace{-0.2cm}\small{
\begin{equation} \label{forwardstop}
    \hat{k}_F=\max\left\{ k\in \{1,\ldots,I\}:-\frac{1}{k}\sum_{i=1}^{k}\log(1-p_i)\leq \alpha\right\}\vspace{-0.2cm}
\end{equation}}\normalsize
where $\left\{p_i\right\}_{i=1}^{I}$ is the sequence of raw \textit{p}-values of the ordered hypotheses. The cutoff $\hat{k}_F$ 
indicates rejection of all hypotheses $H_0^1,\ldots,H_0^{\hat{k}_F}$.

\vspace{0.15cm}$\xrightarrow{}$ In summary, given a sample $x_1,\ldots,\,x_n$ of size $n$ and $\left\{u_i\right\}_{i=1}^I$ a reasonable set of candidate thresholds (as before), the ALGFSM works as follows: 
\begin{enumerate}
    \item For each candidate threshold $u_i$, $i=1,\ldots,I$:
    \small\begin{itemize}
        \item[a)] Compute the GoF measure $z^{GP}_i$ in \eqref{ZGP} according to the described fitting process;
        \item[b)] Compute the corresponding \textit{p}-value $p_{i}$ in \eqref{pvalZGP};
    \end{itemize}\normalsize
   \item Apply the \emph{ForwardStop} stopping rule in \eqref{forwardstop} to the set of $I$ \textit{p}-values of the ordered hypotheses, retrieving the cutoff $\hat{k}_F$;
   \item The lowest threshold after which the tail behaviour of the underlying distribution can be considered approx. GPd is automatically selected as $u^*=u_{\hat{k}_F+1}$ -- the first level above which the corresponding $t_4$ behaves sufficiently closely to the expected behaviour of the GPd's $\tau_4$. No threshold is selected if all $|z^{GP}_i|$ are too large (equivalently, all \textit{p}-values too small).
\end{enumerate}
Performance of this method is, as expected, closer to that of the ALCBSM than of the heuristic ALRSM, since the former was constructed on the same theoretical, asymptotic ground which does not play an explicit part in the latter.

\vspace{0.2cm}\underline{OBS:} Simulation studies (here omitted) regrettably show significant efficiency and accuracy loss of the suggested methodologies, compared to the ALRSM and other state-of-the-art methods, regarding TS and parameter and RL estimation. Thresholds selected tend to be smaller than the appropriate level, and estimation suffers from both considerable bias and uncertainty. Also, computational intensity of both processes makes them unsuitable for large scale data batches requiring simultaneous analysis. As such, the asymptotically justified ALCBSM and ALGFSM cede superiority to the heuristic ALRSM.

\section{Significant Wave Heights Data Sets}
The hindcasts of storm peak significant wave heights (SWH) data sets used to illustrate the proposed methodologies are available and were previously studied in \cite{lombaalves2020} and \cite{northropetal2017} -- refer to these works for a full description of the data: 
315 SWH registered in the Gulf of Mexico (GoM) from September 1900 to September 2005, averaging 3 yearly obs., and 628 SWH registered in October through March, from 1964 to 1995 in the North Sea (NS), averaging $\approx 20.26$ yearly records. These previous studies suggest adequacy of a heavy tailed GPd fit to the GoM data, unlike the bounded tailed GPd judged more suitable for the NS data. Also, there was strong indication that a threshold above the 70\% sample quantile would be most appropriate for both sets, giving shape parameter estimates coherent with the expected tail weights from the preliminary analysis. The ALRSM results in \cite{lombaalves2020} were concordant with this assessment.

Table \ref{tab:GOM_NS_asymp} summarizes the GPd-POT analysis performed, aiming at comparison of threshold selection by the three mentioned methods, and drawing inference regarding 100 and 10 000 year RL (in meters). Selections through the ALCBSM and ALGFSM follow the procedures in Section 2, while results for the competing methodology ALRSM were drawn from \cite{lombaalves2020} (where results from other literature suggestions are also shown for these series). \vspace{-0.1cm}

\begin{table}[h!]
  \centering\tiny{
    \begin{tabular}{ccccccccc}
    &&&\emph{Sample}&&&&&\\
    \multicolumn{1}{c}{\emph{Data}} & \multicolumn{1}{c}{\emph{Method}} & \multicolumn{1}{c}{$I$} & \multicolumn{1}{c}{\emph{quantile-\%}} & \multicolumn{1}{c}{$u^*$} & \multicolumn{1}{c}{$n^*$} & \multicolumn{1}{c}{$\hat{\xi}$} & \multicolumn{1}{c}{$\widehat{RL_{100}}$} &\multicolumn{1}{c}{$\widehat{RL_{10000}}$}  \\
    \toprule
    \multicolumn{1}{c}{\multirow{6}[10]{0.6cm}{\textbf{\textit{Gulf of Mexico}}}} & \multicolumn{1}{c}{\multirow{2}[2]{*}{\textbf{ALCBSM}}} & \textbf{I=10}  & 47.5  & 2.578 & 165   & \textit{-0.195} & \textit{11.11}  & \textit{14.88} \\
\cmidrule{3-9}          &       & \textbf{I=20}  & 50.9  & 2.859 & 155   & \textit{-0.064} & \textit{12.29} &  \textit{21.25} \\
\cmidrule{2-9}\morecmidrules\cmidrule{2-9} & \multicolumn{1}{c}{\multirow{2}[2]{*}{\textbf{ALGFSM}}} & \textbf{I=10}  & 25    & 1.660 & 236   & \textit{-0.183} & \textit{11.14} & \textit{15.15} \\
\cmidrule{3-9}          &       & \textbf{I=20}  & 25    & 1.660 & 236   & \textit{-0.183} & \textit{11.14} &  \textit{15.15} \\
\cmidrule{2-9}\morecmidrules\cmidrule{2-9}         & \multicolumn{1}{c}{\multirow{2}[2]{*}{\textbf{ALRSM}}} & \textbf{I=10}  & 70    & 3.976 & 95    & 0.146 & 14.40 & 35.18 \\
\cmidrule{3-9}          &       & \textbf{I=20}  & 73.1  & 4.182 & 85    & 0.173 & 14.65 & 38.58 \\
    \toprule
    \multicolumn{1}{c}{\multirow{6}[10]{0.6cm}{\textbf{\textit{North Sea}}}} & \multicolumn{1}{c}{\multirow{2}[2]{*}{\textbf{ALCBSM}}} & \textbf{I=10}  & 25    & 2.204 & 470   & \textit{-0.244} & \textit{11.19}  & \textit{12.41 }\\
\cmidrule{3-9}          &       & \textbf{I=20}  & 25    & 2.204 & 470   & \textit{-0.244} & \textit{11.19} &  \textit{12.41} \\
\cmidrule{2-9}\morecmidrules\cmidrule{2-9}         & \multicolumn{1}{c}{\multirow{2}[2]{*}{\textbf{ALGFSM}}} & \textbf{I=10}  & 25    & 2.204 & 470   & \textit{-0.244} & \textit{11.19}  & \textit{12.41} \\
\cmidrule{3-9}          &       & \textbf{I=20}  & 25    & 2.204 & 470   & \textit{-0.244} & \textit{11.19} &  \textit{12.41} \\
\cmidrule{2-9}\morecmidrules\cmidrule{2-9}         & \multicolumn{1}{c}{\multirow{2}[2]{*}{\textbf{ALRSM}}} & \textbf{I=10}  & 77.5  & 4.809 & 142   & -0.346 & 10.72  & 11.37 \\
\cmidrule{3-9}          &       & \textbf{I=20}  & 80.5  & 5.113 & 123   & -0.355 & 10.71 & 11.33 \\
   \bottomrule
    \end{tabular}%
  }\caption{TS and inference for the GoM and NS series by three TS methods; PWM estimates in italic, ML estimates in straight font (for the ALRSM, plugging in the PWM estimates yields similar results).}\label{tab:GOM_NS_asymp}
\end{table}\vspace{-0.2cm}

\normalsize 
There are clear discrepancies between results from the asymptotically-based methods and the ALRSM:\small
\begin{itemize}
    \item Selected levels are considerably lower than those from the ALRSM (which was expected given the simulation studies conducted) -- the lowest candidate is frequently selected;
    
    \item PWM estimates of $\xi$ yielded by the ALCBSM and ALGFSM for the GoM data are unsatisfactory, as positive estimates were expected from sensible analysis of preliminary plots; this directly translates into very low estimated RL, which is problematic for risk analysis;
    
    \item For the NS, both methodologies produce equal results -- larger $\hat{\xi}$ and consequent RL also a little higher than those of the ALRSM (not as significant as for the GoM data given lightness of the tail).
\end{itemize}

\normalsize
The ALCBSM and ALGFSM take around 1 or 2 seconds to produce the full inference, which is not prohibiting of their use here, as happens for simultaneous analysis of large sample batches. As foreseen, results from the newly introduced methods are not satisfactory for these sets, compared to the also L-moment based ALRSM.

\normalsize For illustration, we show for the NS data and $I=10$: the GPd LMRD with $(\mathnormal{t}_{3,u_i},\mathnormal{t}_{4,u_i})$, with 95\% CI computed as \eqref{ICt4dadot3} and \eqref{ICt3dadot4}, for application of the ALCBSM -- Figure \ref{fig:ALCBSM}; the $Z^{GP}_i$ GoF measure in \eqref{ZGP}, corresponding raw \textit{p}-values \eqref{pvalZGP} and adjusted \emph{ForwardStop} values for \eqref{forwardstop}, for application of the ALGFSM -- Table \ref{tab:ALGFSM}.

\vspace{0.3cm}
  \begin{minipage}{0.45\textwidth}
  \centering
      \includegraphics[width=0.85\textwidth]{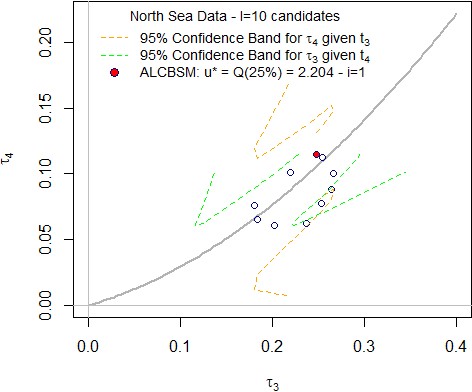}
      \captionof{figure}{ALCBSM: NS data, I=10 candidates}
      \label{fig:ALCBSM}
  \end{minipage}
  \hfill
  \begin{minipage}{0.47\textwidth}
   \tiny{
    \hspace{-0.4cm}
    \begin{tabular}{c||cccc}
\multicolumn{1}{c||}{\multirow{2}{*}{$i$}} & \multirow{2}{*}{$u_i$} & \multirow{2}{*}{$Z^{GP}_i$} & $raw$ & $Forward$\\
&  & &$p_i$& $Stop$\\
\toprule
           1     & 2.204 & -0.559 & 0.576 & 0.858 \\
           2     & 2.444 & -0.219 & 0.826 & 1.304 \\
           3     & 2.800 & 0.992 & 0.321 & 0.999 \\
           4     & 3.193 & 1.596 & 0.110 & 0.778 \\
           5     & \textbf{3.490} & \textbf{1.663} & \textbf{0.096} & \textbf{0.643} \\
           6     & \textbf{3.822} & \textbf{2.057} & \textbf{0.040} & \textbf{0.542} \\
           7     & 4.246 & 0.936 & 0.349 & 0.526 \\
           8     & 4.809 & 0.143 & 0.887 & 0.733 \\
           9     & 5.697 & -0.341 & 0.733 & 0.798 \\
           10    & 7.019 & -0.306 & 0.759 & 0.861 \\
          \bottomrule
    \end{tabular}}
  \vspace{0.51cm}\hspace{0.35cm}\captionof{table}{ALGFSM: NS data, I=10 candidates}\label{tab:ALGFSM}
  \end{minipage}\\

\normalsize This closer analysis shows the automatic procedures ignore some issues, such as the possibility of threshold acceptance conditions, while satisfied at lower levels, being violated for somewhat higher candidates -- subjective analysis of Figure \ref{fig:ALCBSM} and Table \ref{tab:ALGFSM} can suggest the alternative choice $u^*=u_{7}$, closer to the ALRSM's selection.

\vspace{0.2cm} \underline{General Conclusions:} These poor results compared to ALRSM (under suitability of the POT-GPd), are due to two main factors: less strict selection criteria naturally lead to lower chosen levels, and more complex procedures (e.g. simulation of samples from fitted Kappa) are typically more computationally demanding. We conclude that the proposed ALCBSM and ALGFSM, while having some usefulness as validation techniques for detection of deviations from expected GPd behaviour of excesses' samples, are not appropriate as stand-alone threshold selection methods for Extreme Value Analysis.

\section*{Acknowledgments}
This work is partially financed by national funds through FCT –- Fundação para a Ciência e a Tecnologia under project UIDB/00006/ 2020 (JSL \& MIFA) and PhD grant SFRH/BD/130764/2017 (JSL). 

\bibliografia{
\bibitem{baderetal2018}
Bader, B., Yan, J., Zhang, X. (2018). Automated threshold selection for extreme value analysis via ordered goodness-of-fit tests with adjustment for false discovery rate. \textit{Ann Statist} 12, 310--329.

\bibitem{BalkDeHaan1974} 
Balkema, A., de Haan, L. (1974). Residual life time at great age. \textit{Ann Probab} 2, 792--804.

\bibitem{hosking1986}
Hosking, J.R.M. (1986). \textit{The theory of probability weighted moments.} Research Report RC12210, IBM Corporation, New York.

\bibitem{hosking1994}
Hosking, J. R. M. (1994). The four-parameter kappa distribution. \textit{IBM J Res Dev} 38(3), 251--258.

\bibitem{hoskingwallis1997}
Hosking, J.R.M., Wallis, J.R. (1997). \textit{Regional Frequency Analysis -- An Approach Based on L-moments}. Cambridge University Press, Cambridge.

\bibitem{langousis2016}
Langousis, A., Mamalakis, A., Puliga, M., Deidda, R. (2016). Threshold detection for the generalized Pareto distribution: Review of representative methods and application to the NOAA NCDC daily rainfall database. \textit{Water Resour Res} 52, 2659--2681.

\bibitem{northropetal2017}
Northrop, P. J., Attalides, N., Jonathan, P. (2017). Cross-validatory extreme value threshold selection and uncertainty with application to ocean storm severity. \textit{J R Stat Soc C} 66, 93--120.

\bibitem{pickands1975} 
Pickands, J. (1975). Statistical inference using extreme order statistics. \textit{Ann Statist} 3, 119--131.

\bibitem{scarrottmacdonald2012}
Scarrott, C., MacDonald, A. (2012). A review of extreme value threshold estimation and uncertainty quantification. \textit{REVSTAT -- STAT J} 10, 33--60.

\bibitem{lombaalves2020}
Silva Lomba, J., Fraga Alves, M. I. (2020). L-moments for automatic threshold selection in extreme value analysis. \textit{Stoch Environ Res Risk Assess} 34, 465--491.



}
\end{document}